\documentclass[a4paper,12pt]{article}
\usepackage{bm}
\usepackage{amsmath}
\usepackage{amssymb}
\usepackage[dvips]{graphicx}
\usepackage{pdflscape}
\usepackage{multirow}
\usepackage[sort]{natbib}
\usepackage{soul}
\usepackage{xr}
\usepackage{amsmath}
\usepackage{graphicx}
\usepackage{enumerate}
\usepackage{natbib}
\usepackage[hyphens]{url} % not crucial - just used below for the URL 
\usepackage[utf8]{inputenc}
\usepackage{mathtools}
\usepackage{amsbsy}
\usepackage{amssymb}
\usepackage{color}
\usepackage{ulem}
\usepackage{placeins}
\usepackage{booktabs}
\usepackage{multirow}
\usepackage{hyperref}
 \usepackage{booktabs,subcaption,amsfonts,dcolumn}
\usepackage{xcolor,colortbl}
\usepackage{subcaption}
\usepackage{verbatim}
\usepackage{chngcntr}
\usepackage{amsthm}

\definecolor{Gray}{gray}{0.85}
\definecolor{LightCyan}{rgb}{0.88,1,1}
 
\newcolumntype{b}{>{\columncolor{Gray}}c}
\newcolumntype{a}{>{\columncolor{red}}c}
\newcolumntype{g}{>{\columncolor{green}}c}

\title{Fisher's Legacy of  Directional Statistics, and Beyond to Statistics  on Manifolds \footnote{\bf Based on  ``The  40th  Fisher Memorial Lecture'' delivered  in November 2022, Oxford. For the link to the talk, visit\ $$http://www.senns.uk/FisherWeb.html $$ The first topic of the lecture is covered in \citet{mardia2023a} and this is the second topic.}}

\author{ Kanti V. Mardia\\
University of Leeds and University of Oxford.}
\begin{document}
  \date{} % USE THIS IF NO DATE IS TO BE USED

 \maketitle
\begin{abstract}

It will not be an exaggeration to say that R A Fisher is the Albert Einstein of Statistics. He pioneered almost all the main branches of statistics, but it is not as well known that he opened the area of Directional Statistics with his  1953 paper introducing  a distribution on the sphere which is now known as the Fisher distribution. He stressed that for spherical data one should take into account that the data is on a manifold. We will describe this Fisher distribution  and reanalyse his geological data. We also  comment on  the  two goals he  set himself in that paper,  and how he reinvented   the von Mises distribution on the circle.  Since then, many extensions of this distribution have appeared bearing Fisher's name such as the von Mises  Fisher distribution and the matrix Fisher distribution. In fact, the subject of Directional Statistics has grown tremendously in the last two decades with new applications emerging in Life Sciences, Image Analysis, Machine Learning and so on. We give a  recent new method of constructing the  Fisher type distribution which has been motivated by some problems in Machine Learning.  The subject related to his distribution has evolved since then more broadly as Statistics on Manifolds which also includes the new field of Shape Analysis. We end with a historical note pointing out some correspondence between D'Arcy Thompson  and R A Fisher related to  Shape Analysis.

\end{abstract}

AMS (2000) subject classification. Primary 62H15, 62R30; Secondary 62E10,
62F03.\\


\section {Introduction}

 Fisher  opened the area of Directional Statistics with his pioneering 1953 paper, \citet{fisher1953},  introducing what is now known as the Fisher distribution on the sphere. He stressed that for spherical data one should take into account that the data is on a manifold. Since then, many extensions of this distribution have appeared bearing Fisher's name as we will describe. In fact, the subject of Directional Statistics has grown tremendously in the last two decades with new applications emerging in Life Sciences, Image Analysis, Machine Learning, and so on.

One of the features of Fisher's work is that the starting point was often a motivating application from scientists and applied statisticians. For Directional Statistics, this was the problem of pole reversal raised by the geologists Mr J. Hospers and Professor S. K. Runcorn.

  In Sections  \ref{DDA main1} and   \ref{DDA main2}, we give   an overview of the subject 
 and   discuss in particular  the extension of the Fisher distribution on the sphere to the  hypersphere, which is now known as the von Mises-Fisher distribution, examined in Section \ref{VmF}.
   In  Section \ref{FisherJustify}, we describe, some features of  the seminal paper of Fisher (1953) and his reasoning  on why linear statistics is not meaningful  in the context of his practical applications. Also we describe  how this work was taken up by Geoff Watson (starting with \citet{Wat1956} which was  written in May, 1955) who made it more accessible and popular. Historically,  it is important to learn that though Fisher's  paper appeared in 1953 but  he  already formed these  ideas  in the 1920's. It was the the geological application that made him  go back and  activate this work (Section  \ref{FisherWhen}).  We reanalyse his geological data in Section \ref{HospersData}.
   
  In general, the maximum likelihood methods for directional distributions are not computationally straightforward. A  new approach,  the score matching estimate, will be presented for the von Mises Fisher distribution in Section \ref{estimates}. However, in his paper, Fisher produced estimators using the distribution of the ``sufficient statistics `` and we examine closely his approach in Section  \ref{FisherEst}.  He dealt with  two cases,  known pole and  known axis. In the second case, his estimation method is somewhat questionable and has  not been used---even he did not use it  in his own paper of 1953.  It is not commonly known  that Fisher derived the von Mises distribution independently; his aim was  to promote his Fiducial argument (see Section \ref{FisherOnVM}). He mentioned that his paper  has two goals and we assess  these   in  Section \ref{Goals}. 
  
  The subject of Statistics on Manifold is still evolving, including  a  new method to construct the Fisher type  matrix distribution which has been motivated by some problems in Machine Learning. We examine this new approach in Section \ref{wrap}. Finally, in Section  \ref{shape}, we give a historical note pointing out some correspondence between D'Arcy Thompson (the pioneer of Shape Analysis) and R A Fisher where we could have gained Fisher's insight into Shape Analysis; however,  this collaborative work did not happen.

\section {Statistics on Manifolds/Directional Statistics} \label{DDA main1}

Big data, high dimensional data, and sparse data are all  new frontiers of statistics. Changing technologies have created this flood of data with new challenges, and have led to a substantial  need for new modelling strategies and data analysis. 
 There are  data which are essentially  not Euclidean - - the data sit on a manifold.  Even for the   simplest non-Euclidean manifold, the circle, with angular data, using the arithmetic average cannot make sense, as is well known.  Consider the average of two
angles $1^\circ$ and  $359^\circ$ is  
$$\frac{1^\circ+ 359^\circ}{2}= 180^\circ \; ! $$
Of course, it should be $0^\circ .  $ 
 That is,   the non-Euclidean setting throws up many major challenges, both mathematical and statistical and so more care is needed. 
In simple terms, Statistics on Manifolds deals with  non-Euclidean variables
 driven mainly by the  underlying geometry. Examples include  
  circle, sphere, torus,
   and  shape spaces.  
   
   The subject of Directional Statistics has grown tremendously, especially since the 1980's, with advances in  ``Statistics on Manifolds" leading to new distributions on the hyper-sphere, torus, Stiefel manifold, Grassmann manifold and so on. The progress in this area  can be seen through several books published since then: Nick  \citet{fisheretal1987}, Nick \citet{fisher1993}, \citet{mardiajupp2000},   \citet{jammala_seng2001}, \citet{ley_verdebout2017} and \citet{ley_verdebout2018}. There has been a  recent special issue of Sankhy$\bar a$   edited by \citet{spissue2019}.  Further, \citet{pewsey_eduardo2020} have given a comprehensive survey of directional statistics and in the discussion to the paper, \citet{mardia2021} has given a brief history of the subject. In particular, to note that the methods of   Principal Component Analysis on the torus are now well established (see, for example, \citet {mardia2021principal}, \citet {kvm2023a}). Another major development is the new  area of discrete dsitributions in directional statistics, see \cite{mardiasriram2020ar} and  \citet{mardiasriram2023}; these two papers take forward Karl Pearson's challenges  on his roulette wheel  data  of 1890's,  though this did not draw attention of Fisher.  
   
\subsection { Fisher 1953's  paper: A landmark paper in  Directional Statistics, and the role of Geoff Watson} \label{FisherJustify}
 The paper of  R. A. Fisher entitled  ``Dispersion on a sphere" which appeared in 1953, motivated by   remanent magnetism data. 
  He begins with justifying the  need for such work, as the following  quotes from the paper shows  (bold sentences and bullet points   for  emphasis.)
\begin{itemize}
\item `` The theory of errors was developed by Gauss primarily in relation to the needs of
astronomers and surveyors, making rather accurate angular measurements. Because
of this accuracy it was appropriate to develop the theory in relation to an infinite
linear continuum, or, as multivariate errors came into view, to a Euclidean space of
the required dimensionality. The actual topological framework of such measurements,
the surface of a sphere, is ignored in the theory as developed, with a certain 
gain in simplicity.
\item It is, therefore, of some little mathematical interest to consider how the theory
would have had to be developed if the observations under discussion had in fact involved errors so large that {\bf  the actual topology had had to be taken into account.}
\item The question is not, however, entirely academic, for there are in nature vectors with
such large natural dispersions. The remanent magnetism found in igneous and
sedimentary rocks,..., do show such considerable
dispersion that an adequate theory for the combination of such observations is now needed. 
\item My examples are drawn from the very fine body of data on the remanent magnetism
of Icelandic lava flows, historic and prehistoric, put at my disposal by Mr J. Hospers...''
\end{itemize}
In  fact, this subject took off  when the 1953 paper came to the attention  of  Geoff Watson, one of the pioneers of Directional Statistics.  
It is interesting to record how this happened.
  In his conversation  with Rudy Beran and Nick Fisher(\citet{beran1998}), Watson says 
  
  {\bf ``We sent a food parcel to his daughter, Joan, in England, at his request. As a ``thank you" note he sent me a reprint of his 1953 paper  which is not easy reading ... Anyhow I had a look at it and suddenly saw that I could clarify things." } 
 (This is referring to Joan Box, a daughter of Fisher who wrote his biography, \citet{box1978} .)
 
\begin{figure}[!htb]
\caption{\label{FisherPhoto}Fisher 1953 in a Picnic in Australia; {\bf   Photo  by Geoff Watson --  Gifted  this version to the author  in 1990. } }
\begin{center}
\includegraphics[width= 8cm,angle=0]{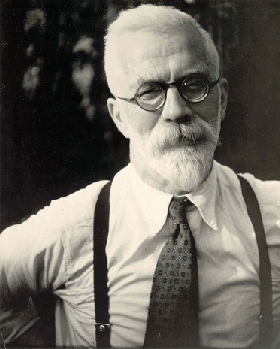}
\end{center}
\end{figure}

 We want to recall that  Fisher visited Melbourne University  while Watson was Senior Lecturer in Statistics from 1951  and an acting head of the department; Watson looked after him during his visit  \citet{Wat1990}.  In fact,  Figure \ref{FisherPhoto} shows 
 one of the   Fisher's photos taken during this period by Watson in a picnic; this was given to the author personally by Geoff.

 Following that, Geoff Watson  and  Michael Stephen made several key contributions in 1956- 1970
 (starting from Watson's first paper of  1956, \citet{Wat1956}, written in May, 1955.)
In    1970-1980  the subject came into the limelight, 
partly with my  1972 directional book \citet{mardia1972}  but perhaps 
  more so with my  discussion paper in the Journal  of the Royal Statistical Society  which included many eminent discussants, David Cox, Henry Daniel, David Kendall, John Kingman among others. There was   something of  a lull in the subject  during 1990 to 1999, though a few books came out, see \citet{mardia2021} for more details. 
From   2000,   a resurgence of interest led to many new advances, mainly because of  the recognition by Image Analysts and Life Scientists of its importance. This momentum now continues as most of   statisticians do regard Directional Statistics as a mainstream statistical topic.

\subsection {When was the material of the  1953 paper written?}\label{FisherWhen}
There has been some discussion on  why and  when was the material written for the 1953 paper. We first quote \citet{box1978} p.439 on  the roles of Fisher, Hospers and Runcorn (Hospers was Runcorn's student who collected the data) : 
\begin{itemize}
\item   
Runcorn  was at once struck by the fact that these lavas  grouped themselves into those along the present field and those
opposite, in a manner strongly suggesting that {\bf the direction of magnetization
of the earth had been reversed at some periods of its history.}
\item  
Runcorn explained these results to Fisher and asked him how the particular
statistical problems of testing homogeneity should be solved.
 Fisher, pulling out some old notes he had made before 1930  in connection with his paper on
tests of significance in harmonic analysis [1929], set to work to apply the method he had then devised to the data now in hand .
\item  
  Hospers quickly used the method  in three very important papers  which laid
the basis for the paleomagnetic work in the Tertiary Age.
\end{itemize}
Hospers used Fisher's 1953 paper's preprint for his three papers described above. In Section \ref{HospersData}, we give full details of  Hospers' data used by Fisher.

 Another version on this point  is given  as follows in Nick \citet{fisheretal1987}, pp. 12-13, quoting George Barnard (letter to Nick Fisher, 30 June 1981.) 
\begin{itemize}
\item 
Fisher sent me an offprint of his paper ``Dispersion on a sphere"  when it came out,  $\ldots $ while I saw you could use the length of the vector sum to test isotropy, I had not seen how to do estimation. 
He replied that it was easier for him, since he had, in the twenties, asked
himself what would be the analogue, on the sphere, of the normal density in the plane, and had made some notes on it. 
\item  
When approached by (I presume) Hospers he was able to go to his filing cabinet and pull out the notes, and answer the question on the spot .... the figure 1922 comes into my head, for Fisher's first work.
 This seems rather early; but he was in touch with Eddington, the astronomer, in 1920, and may well have gone on thinking about the problems he had raised.
\end{itemize}

Another account of this paper has been given by  Persi  Diaconis in 1988 (\citet{Persi1988}, p. 171) 
which we now quote.
\begin{itemize}
\item
I cannot resist reporting some background on Fisher's motivation for working with the distribution discussed above. This story was told to me in 1984 by the geologist Colin B. B. Bull. Dr. Bull was a student in Cambridge in the early 1950's….. Fisher asked what area Bull worked in. Bull explained that a group of geologists was trying to test Wegener's theory of continental drift. Wegener had postulated that our current continents used to nest together. 
\item
They searched for data that were closer to geology. They had hit on the distribution of magnetization angle in rocks. This gave points naturally distributed on the sphere. They had two distributions (from matching points on
two continents) and wanted to test if the distributions were the same.
\item
Fisher took a surprisingly keen interest in the problem and set out  to learn the relevant geology. In addition to writing his famous paper (which showed the distributions were different) he gave a series of talks at the geology department to make sure he'd got it right. 
\item
Why did Fisher take such a keen interest? A large part of the answer may lie in Fisher's ongoing war with Harold Jeffries. They had been rudely battling for at least 30 years over the foundations of statistics. Jeffries has never really accepted (as of 1987!) continental drift. It is scarcely mentioned in Jeffries' book on geophysics. 
\end{itemize}

These all accounts point to the fact that Fisher has developed the distribution around 1922 and his subsequent contacts with geologists in the early 1950's led to his 1953's paper.

\section{The  Fisher Distribution and other  Directional Distributions}\label{DDA main2}
We first recall some manifolds before giving some relevant directional distributions.
\\
{\bf Sphere.}  
The sphere
\begin{equation}
S_p = \{\boldsymbol{x} \in \mathbb{R}^{q}: \ \boldsymbol{x}^T\boldsymbol{x} = 1\}, \ q=p+1,
\end{equation}
  represents the
space of unit vectors or ``directions'' in $\mathbb{R}^q$.  We have the circle  when $p=1$.
\\
{\bf Real projective space.}
The  real projective space consists of the ``axes'' or ``unsigned directions'' $\pm \boldsymbol{x}$.  In some
sense this space is half of a sphere; it can also be represented as
the space of rank 1 projection matrices,
\begin{equation}
\label{eq:rpp}
 \mathbb{R}P_p
 = \{\boldsymbol{P} \in \mathbb{R}^{q\times q}: 
\boldsymbol{P}=\boldsymbol{P}^T, \; \boldsymbol{P}^2=\boldsymbol{P},  \ trace \; \boldsymbol{P}=1 \}.
\end{equation}  
A rank one projection matrix can be written as $\boldsymbol{P}=\boldsymbol{x}\boldsymbol{x}^T$ where $\boldsymbol{x}$ is a unit
vector.  
\\
{\bf Rotation matrices.}
The special orthogonal group of $r \times r$ rotation
matrices is defined by
\begin{equation}
SO(r) = \left\{\boldsymbol{X} \in \mathbb{R}^{r \times r}: \ 
\det{\boldsymbol{X}}=1, \ \boldsymbol{X}^T\boldsymbol{X} = \boldsymbol{I}_r \right\}.
\end{equation}

On each of these spaces, there is a unique uniform distribution which
is invariant under rotations.  Further each of these spaces is
naturally embedded in a Euclidean space.  A natural
``linear-exponential'' family of distributions can be generated by
letting the density (with respect to the uniform measure) be
proportional to the exponential of a linear function of the embedded
variables.  We now list some distributions with their names and space as listed  in Table \ref{listDist}.
\begin{itemize}
\item The von Mises-Fisher distribution (with respect to Lebesgue measure)   on $S_p$, parameter $\boldsymbol{\alpha} \in \mathbb{R}^q, \ q=p+1$:
$$
f(\boldsymbol{x}) \propto \exp(\boldsymbol{\alpha}^T\boldsymbol{x}), \quad \boldsymbol{\alpha},\boldsymbol{x} \in \mathbb{R}^q, \boldsymbol{x}^T\boldsymbol{x}=1,
$$
where usually we write $ \boldsymbol{\alpha}=\kappa  \boldsymbol{\mu}$ with  $ \boldsymbol{\mu}^T \boldsymbol{\mu}=1.$
The distribution   is analogous to (for the concentrated case) a $p$-dimensional isotropic normal distribution. 
More details on this distribution are given in Section \ref{VmF}.
\item The Bingham distribution  (with respect to Lebesgue measure)  on $S_p$, symmetric parameter matrix $\boldsymbol{A} (q
  \times q)$; $\boldsymbol{A}$ and $\boldsymbol{A}+ \lambda \boldsymbol{I}_q$ define the same distribution:
$$
f(\boldsymbol{x}) \propto \exp(-\boldsymbol{x}^T \boldsymbol{A} \boldsymbol{x}) = \exp\{ - trace(\boldsymbol{A}\boldsymbol{x} \boldsymbol{x}^T)\}, \quad  \boldsymbol{x}\in
\mathbb{R}^q,\quad  \boldsymbol{x}^T  \boldsymbol{x}=1,
$$
that is, quadratic-exponential on $S_p$ and  linear-exponential on $\mathbb{R}P_p$.
The distribution   is analogous to (for the concentrated case) a $p$-dimensional general normal distribution.\\

\end{itemize}

\begin{itemize}

\item The Matrix Fisher  distribution on $SO(r)$, parameter matrix $\boldsymbol{F} (r \times r)$,
$$
f(\boldsymbol{X})  \propto \left\{  \ trace \left(\boldsymbol{F}^T\boldsymbol{X} \right) \right\}, \quad \boldsymbol{X} \in SO(r),
$$
with respect to the underlying invariant ``Haar'' measure.  It is unimodal
about a fixed rotation matrix determined by $\boldsymbol{F}$.
This distribution   is analogous to  (for the concentrated case) a $r(r-1)/2$-dimensional general normal 
distribution.\\

\end{itemize}

\begin{table}[t!]
\caption{\bf  Some common   directional distributions in exponential family: spaces and names.}
\begin{tabular}{lll}
Space & Notation\ \ \ \ & Distributions\\ \hline
circle   & $S_1$      & von Mises ($p=1$) \\  % wrapped Cauchy 
sphere   & $S_p$      & Fisher ($p=2$)\\
         &            & von Mises-Fisher ($p \geq 1$),\\
         &            & Fisher-Bingham\\
real projective space & $\mathbb{R}P_p$ & Bingham\\
         %&            & angular central Gaussian\\
special orthogonal group & $SO(r)$ & matrix Fisher \\ \hline
\end{tabular}
\label{listDist}
\end{table}
%%%%%%%%%%%%%%%%%
For further details on these distributions, see, for example, \citet{mardiajupp2000}. 

%%%%%%
\section{The von Mises-Fisher distribution} \label {VmF}
The  von Mises-Fisher  distribution on $S_p$ has the density with respect to Lebesgue measure  
\begin{equation}
\label{vMFpdf}
f(\boldsymbol{x}) = c_p(\kappa)  \exp(\kappa  \boldsymbol{\mu}^T \boldsymbol{x}), \quad   \boldsymbol{x} \in \mathbb{R}^{p+1},\;\kappa \geq 0,\; \boldsymbol{\mu}^T \boldsymbol{\mu}=1, \ \boldsymbol{x}^T\boldsymbol{x}=1,
\end{equation}
 where 
\begin{equation}
\label{vMFnc}
c_p(\kappa) = 
\frac{ \kappa ^{(p-1)/2 }}  {  { (2\pi)}^{(p+1)/2 } I_{(p - 1)/2}(\kappa)},
\end{equation}
and  $I_{\nu}(.)$ is the modified Bessel function of the first kind and order $\nu$. 
 For $p=1$, it is the von Mises  distribution and for $p=2$, the  Fisher distribution.
 The distribution   has a mode at the mean  direction $\boldsymbol{\mu}$   and   for large concentration parameter  $\kappa$, it has  a $p$-dimensional isotropic normal distribution. 
 For $\kappa=0 $, we have the uniform distribution on the hypersphere.  It is also known as the Langevin distribution since a generalised form was given by  Langevin in 1905 in the context of the theory of magnetism (\citet{Lang1905}).
\\
{\bf The Fisher distribution.} Let $\theta$ denote the  colatitude $ 0 \leq \theta \leq \pi$  and  $\phi$ be the  longitude $0 \leq \phi \leq 2\pi$ in the  spherical polar coordinates.
Fisher (1953) with $p=2$   took    the north pole as the mean direction  ($\boldsymbol{\mu}=(0,0,1)^T$ )  so $\theta$ then represents  the angular displacement from the true mean direction, and  the Fisher  density of $\theta$  in  (\ref {VmF}) simplifies to  
\begin{equation}
\label{Fpdf}
\frac{\kappa}{2 \sinh \kappa} \exp \{ \kappa \cos \theta  \}  \sin \theta ,\quad   0 \leq \theta \leq \pi,    \kappa \geq 0,
\end{equation}
and $\theta$ is independent of $\phi$ which is distributed uniformly on a circle. In general, the distribution has a mode at the mean  direction. Let us write 
 $$ x=\sqrt \kappa \;\theta\; \cos \phi, y= \sqrt \kappa\; \theta\; \sin \phi$$ 
then   for  large $ \kappa$, $(x,y)$ has an  isotropic bivariate  normal distribution with zero means  and unit variances. For $\kappa=0 $, we have the uniform distribution on the sphere.
 \\
%%%%
{\bf Summary statistics on a sphere. }
Let ${\bf  x}_1, \dots, {\bf  x}_n, $ be $n$ points on $S^{p}$. Then the location of 
these
points can be summarised by their sample mean vector in $R^q$, 
which is
\begin{equation}
\label{xbar}
{\bf {\bar x}} = \frac{1}{n} \sum _{i=1}^n{\bf  x} _i.
\end{equation}
Write  the vector $\bar{\bf  x}$ 

\begin{equation}
\label{Rbar}
{\bf {\bar x}} = {\bar R}{\bf {\bar x}}_0,  {\bar R} \ge 0,
\end{equation}
where 
${\bf {\bar x}}_0$ is  the sample {\bf mean direction} and 
 ${\bar R} (= \lVert{\bf {\bar x}}\rVert)$ is the {\bf mean resultant length.}
Note that  $ {\bf {\bar x}}$ is the centre of  gravity with  direction  ${\bf {\bar x}}_0$,  and
${\bar R}$ is its distance from the origin. For the  circular case,  we have  ${\bf {\bar x}}_0^T=\bar R (\cos\bar\theta,\sin\bar\theta)$  where  $\bar\theta$ is the mean  direction.

For further details on the von Mises - Fisher distribution and data analysis, see, for example, \citet{mardiajupp2000}. Next section, deals with some estimators for this distribution.
%%%%%%

\section {Estimation for the von Mises-Fisher distribution.}\label{estimates}
Let  ${\bf  x}_1, \ldots, {\bf  x}_n, $ be  a random sample drawn from the von Mises-Fisher distribution   then the maximum likelihood estimates (MLE) of $\boldsymbol {\mu}, \kappa$
are given by 
\begin{equation}
\label{MLE}
\hat{\boldsymbol {\mu}}_{MLE}=  {\bf {\bar x}}_0; \;
 \hat{\kappa}_{MLE} = A_p ^{-1}(\bar{R}),
 \end{equation}
where $$  A_p(\kappa ) = {I_{(p+1)/2 }(\kappa)}/ {I_{(p - 1)/2}(\kappa)}. $$
Assuming that $\boldsymbol {\mu}$ is known   and writing  in the polar coordinates  with $ \boldsymbol {\mu}^T \boldsymbol {x} = \cos \theta$,  then \citet{mardiaetal2016ar}  have shown that the   Score Matching Estimate (SME)  of $\kappa$ is  simply  
$$\hat{\kappa}_{SME }= p \frac {\sum \cos\theta_i} {\sum \sin^2\theta_i};$$
 there are no  Bessel functions in this expression and there is only  a moderate  loss of efficiency compared to the MLE.

\subsection{Estimation in \citet{fisher1953}} \label {FisherEst}
We now  write the probability density function   of the Fisher distribution in 3D as  
\begin{equation}
f({\bf  x}) = B(\kappa)  \exp(\kappa \boldsymbol{\mu}^T {\bf  x}), \quad   {\bf  x} \in \mathbb{R}^{3},\kappa \geq 0, \boldsymbol{\mu}^T\boldsymbol{\mu}=1,   {\bf  x}^T{\bf  x}=1,
\label{Fisher3D}
\end {equation}
where, for simplicity,  we have written here the constant  $c_2(\kappa)$    in (\ref{vMFnc}) as $B(\kappa)$ which has a simple expression given by 
$$B(\kappa) = \kappa/(2 \sinh \kappa).$$  The pdf is  again with respect to the Lebesgue measure.
\citet{fisher1953} in his Section 2.1 dealt with  {\bf Case 1}: the known pole  case ( i.e. the mean direction $\boldsymbol{\mu}$ is given)  and in his Section 2.2 dealt with {\bf Case 2}:  the case when the known  axis case (i.e.  $\boldsymbol{\mu}$ is given  as  an axes)  to estimate $\kappa$.
 In both these cases, he focused on the distribution of some summary  statistics  to get  an estimate  of  $\kappa$. In {\bf Case 1}, this approach  leads to   the maximum  likelihood estimate (mle)  of  $\kappa$  but not for {\bf Case 2}.  This paper predates what is now known as the Fisher-Neyman factorization theorem and sufficient statistics. In the following, we will use \citet{fisher1953}'s notation wherever possible so the results can be compared with his paper.
 
 \subsubsection{ Case 1: Known pole} 
If  we  know  the true  pole  $\boldsymbol{\mu}$, then the sufficient statistics is 
\begin{equation}
 x = \sum \cos \theta_i = \sum \boldsymbol{\mu}^T {\bf  y}_i, 
 \label{x}
\end{equation}  
where ${\bf  y}_i$ is i-th observed direction vector, $i=1, \ldots, N.$ \citet{fisher1953}   has shown that the distribution of $x$ is 

\begin{equation}
g_N(x) = B(\kappa)^N\exp\{ \kappa x\} P(x,N), 
\label{Pole}
\end{equation}
where 
\begin{equation}
P(x,N) =\frac{1}{(N-1)!}\{ (N-x)^{N-1}  - N(N-2-x)^{N-1}+\ldots -\ldots (-1)^r {N\choose  r}  (N-2r-x) ^{N-1}  \},  
\label{Prob}
\end{equation}
and $r$ is the largest integer less than $\frac{1}{2} (N-x).$
Now can  maximise (\ref{Pole}) with respect to $\kappa$ to get his estimate of $\kappa$, that is, it is the solution of the equation
\begin{equation}
\coth(\kappa) - 1/\kappa = x/N,\quad \kappa>0.
\label{mleCase1}
\end{equation}
This  is,  of course, the mle estimate of $\kappa$  which we would have got by working directly on (\ref{Fisher3D}). Further, we note that the Fisher-Neyman factorization theorem does not need the distribution (\ref{Pole})  of the sufficient statistics for the estimation but  was used by Fisher  rather then going directly to the likelihood.

\subsubsection{Case 2: Known axis} 
\begin{figure}[h]
\begin{center}
\includegraphics[scale=.12,angle=0]{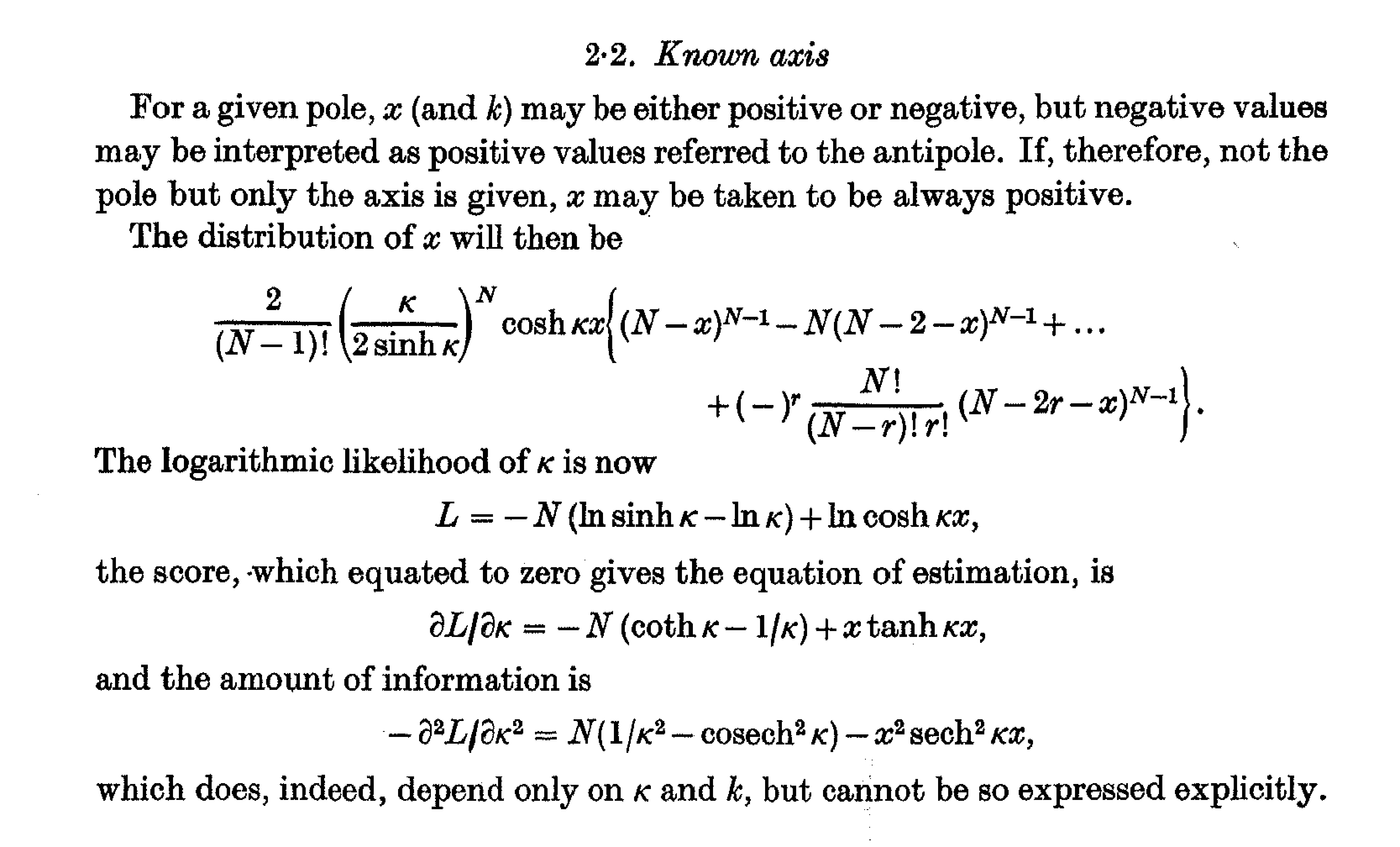}\\
 \end{center}
 \caption{\label{Fsection2.2} Section 2.2 of  \citet{fisher1953} showing his approach  to estimation using the  distribution of the summary statistic $x$  for the known axis case.}
\end{figure}

\citet{fisher1953}  continued to use the  summary statistics $x$ for  known axis given by (\ref{x})  in the form defined in his paper (see, the extract in  Figure \ref {Fsection2.2} of this section in his paper), though  it is no longer a sufficient statistics. 
If all we know is that  the true  axis  $\pm\boldsymbol{\mu}$, then $x = \sum \cos\theta_i = \sum \boldsymbol{\mu}^T {\bf  y}_i$ is not observable.  
  
For  this case, we have to include both $+x$ and $-x$ given  in (\ref{Fisher3D})  finding the probability of $x$ given by (\ref{x}).  
   One approach   is to arbitrarily choose a direction for $\boldsymbol{\mu}$ and   then $x$ is  defined for any $N$. 
When  $ N > 1$, we  must pick a head and a tail for $ \boldsymbol{\mu}$ and  then $x =\sum\boldsymbol{\mu}^T {\bf  y}_i   $  is well defined, but we do not know whether its sign is correct.  Both possibilities $x =\sum \boldsymbol{\mu}^T{\bf  y}_i   $ and $x =-\sum \boldsymbol{\mu}^T {\bf  y}_i   $     need to be included in its distribution,  then  the pdf of $x$ is
\begin{equation}
   g_N(x,\kappa) + g_N(-x, \kappa)=B(\kappa)^N\exp(\kappa x) P(x,N)+B(\kappa)^N\exp(-\kappa x)P(-x,N), 
 \label{Axial All N}
\end{equation}
where $ g_N(.,\kappa) $ is given by (\ref{Pole}).
It can be seen from  (\ref{Prob})  that    $P(x,N) = P(-x,N)$ in (\ref{Axial All N}) leading to (see, Figure \ref {Fsection2.2}) his pdf as 
\begin{equation}
 2  B(\kappa)^N \cosh(\kappa x) P(x,N) .
 \label{Axial All F}
\end{equation}

 However, he uses (\ref{Axial All F})  this sampling distribution to give an estimate of $\kappa$ (see, Figure \ref {Fsection2.2}) so that  it is the solution of the equation
 \begin{equation}
\coth(\kappa) - 1/\kappa = (x/N)\tanh(\kappa x/N).
\label{mleCase2}
\end{equation}
Note that for large $N$, the estimate for {\bf Case 1} given by  (\ref{mleCase1}) is the same as that from  (\ref{mleCase2}).

{\bf The  MLE for the axial case } 
Let $\boldsymbol{\nu}$ be the upper hemisphere representation of the known axis and 
$x =\sum(\boldsymbol{\nu}^T {\bf  y}_i)$. Then $x$ is well defined but we do not know if $x = \sum(\boldsymbol{\mu}^T {\bf  y}_i)$ or $x = -\sum(\boldsymbol{\mu}^T{\bf  y}_i)$ where $\boldsymbol{\mu}$ is the true pole.

Fisher does not give the MLE for this case as his aim again was to use an appropriate summary statistics. We can formulate the estimation problem in this case as follows to get the MLE. There are two unknown parameters, $\kappa \geq 0$ and, say, $\lambda$ with values $-1$ and $+1$ such that $\boldsymbol{\mu} = \lambda\boldsymbol{\nu}$ The parameter $\kappa$ is estimated by the solution to (\ref{mleCase1}) with $x = \sum(\boldsymbol{\nu}^T{\bf  y}_i)$, and $\lambda$ is estimated by $\text{sign}(\sum(\boldsymbol{\nu}^T{\bf  y}_i))$.

 We note that in practice, the case when only the axis of the mean direction is known 
 ({\bf Case 2}) is rare and the estimate  of \citet{fisher1953}  has not been used. Indeed, even in his own  paper he does not give any example for this case.
%%%%%%

\section{Hospers'  remanent magnetism data sets }\label {HospersData}
\begin{figure}
\begin{center}
\includegraphics[scale=.35,angle=0]{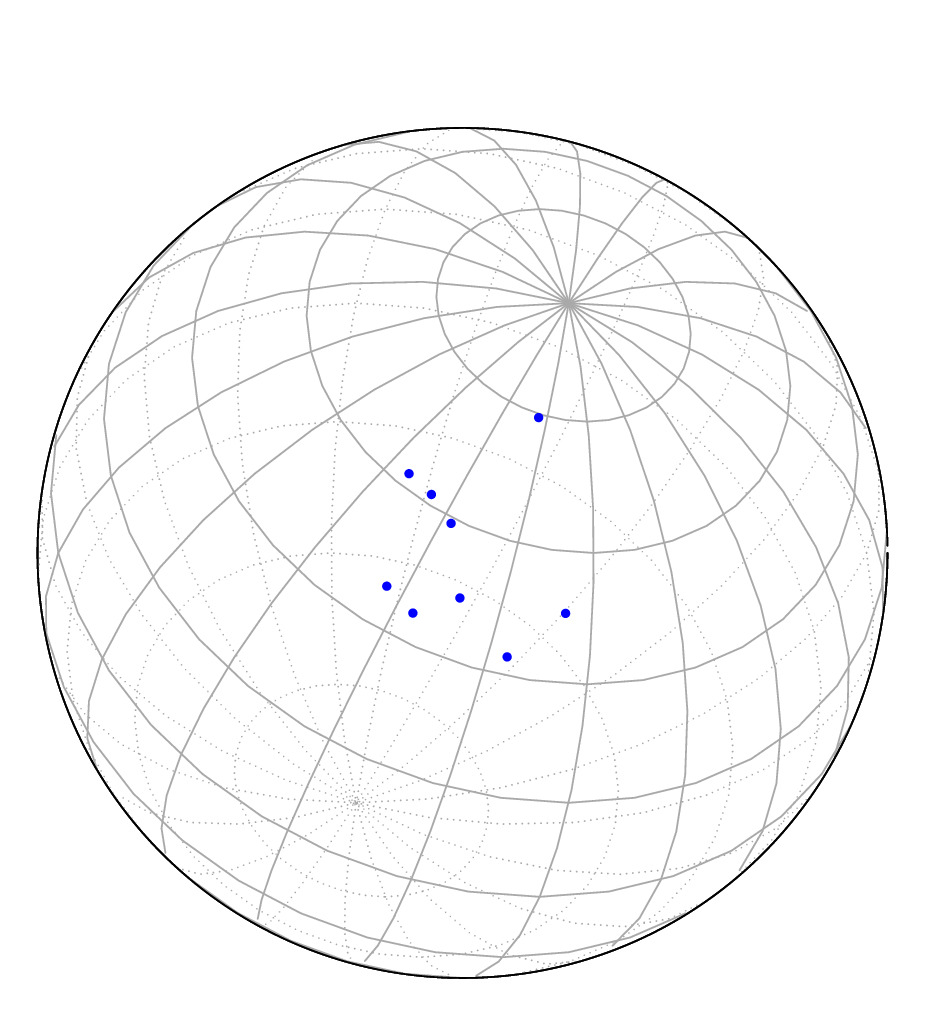}\\
 \end{center}
 \caption{\label{HosperData1Sphere} Hospers` {\bf Remanent magnetism  Data 1}:  Spherical representation.}
\end{figure}

\begin{figure}
\begin{center}
\includegraphics[scale=.35,angle=0]{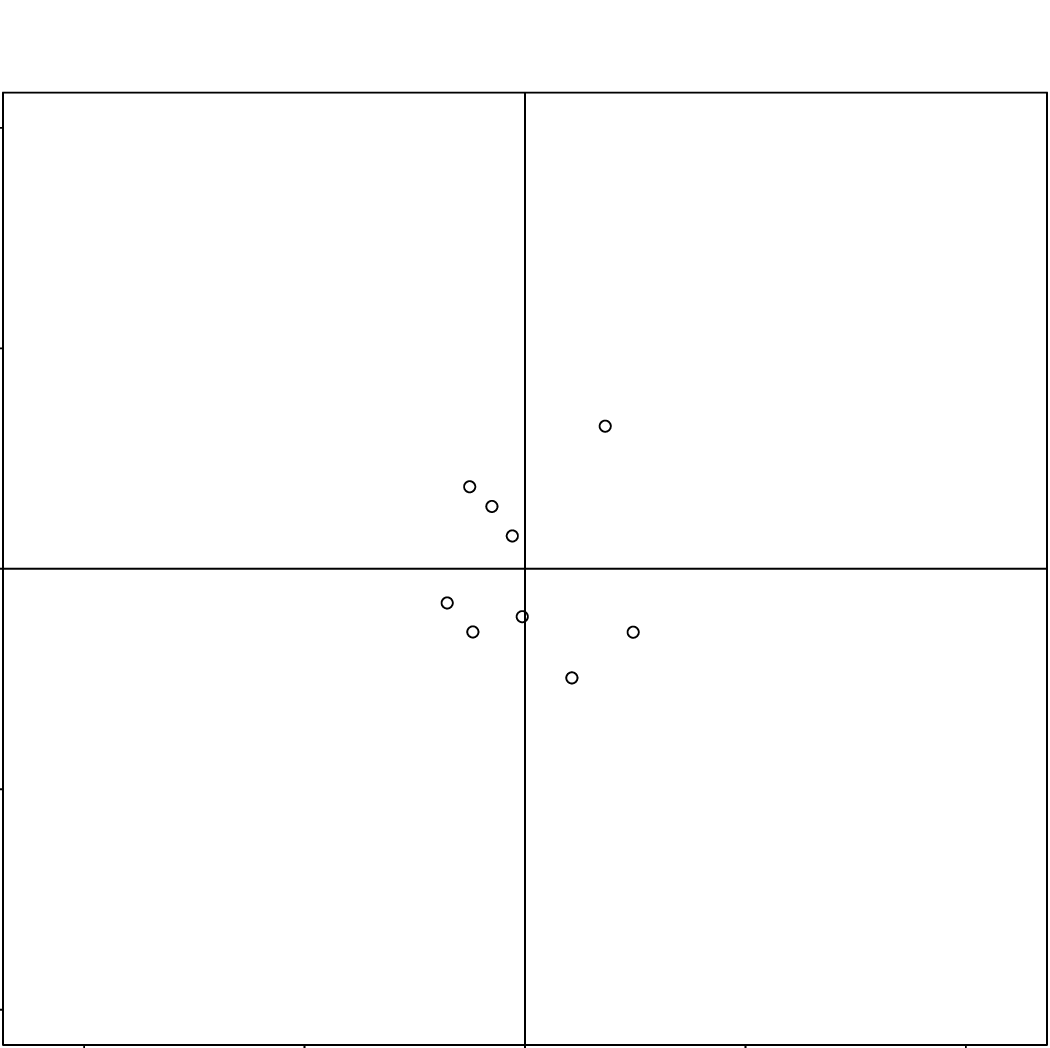}\\
 \end{center}
 \caption{\label{HosperData1Proj} Hospers` {\bf Remanent magnetism Data 1}: The Lambert Equal Area Projection.}
\end{figure}

\citet{fisher1953}  used two Hospers' remanent magnetism data to illustrate his analysis  which  we revisit; we label these as Remanent magnetism  Data 1 and Remanent magnetism Data 2. In fact, Hospers, in a paper published in 1951 in Nature (\citet{Hosper1951})  on remanent magnetism of rocks
 he thanks 
\\ 
 ``Prof. R. A. Fisher for the calculation of the
estimates of precision....”
\\
That is, Hospers and Fisher  had already been
applying the  methodology  to Hospers’ data several years before  Fisher
published his 1953 paper.
\\
{\bf Remanent magnetism Data 1} The data are from the  Iceland lava flow of 1947-1948 with sample size $n=9$ where the full data is given in the paper with the mean direction and  is plotted in Figure \ref{HosperData1Sphere}. 
It is found that  $\hat{\kappa}_{MLE }=39.53 $ so  it is highly concentrated  and we might   have used linear statistics.
 One way is to use the tangent projection at the  mean direction and then    use a linear method. Let $(\theta, \phi) $ be the spherical polar coordinates as used in (\ref{Fpdf}).
It can be shown that the Lambert equal area projection (see, for example, \citet{mardiajupp2000}, p.160) is given by $ (2\sin(\theta/2) \cos\phi, 2\sin(\theta/2) \sin\phi)$ .  Figure \ref{HosperData1Proj} shows the projected data under the projection  and one could  use these values projected values as drawn from a bivariate normal distribution as  $\kappa$ is large. Equivalently,  one can simply take the observed  values of  $(\theta, \phi)$  and assume these are drawn from a bivariate normal distribution and carry out analysis. Fisher must have realised this as a possible option for this data after the analysis,  and his  Remanent magnetism Data 2  described below  in the paper which is not concentrated so it provides a better illustration.
%%%%%%
\\
{\bf Remanent magnetism Data 2.}
  For the  early Quaternary data of Hospers, $n=45$,  Fisher examined the
Hospers's hypothesis  that these observations   are almost diametrically opposite to the simple dipole field (current) which has the mean direction $  (+0.9724, +0.2334,  0)^T. $ 
 That is, we need to  test the reversal hypothesis, and  in the modern terminology, our  null hypothesis is
  $$ H_0 : \boldsymbol{\mu} =  (-0.9724,  -0.2334,  0)^T. $$
From these 45 observations, the sample  mean direction is
$$ \hat{\boldsymbol{\mu}}_{MLE} =  (-0.9545, -0.2978, +0.0172)^T $$ 
which is very close to $\boldsymbol{\mu}$ under $H_0$. Note that  the angle between $\boldsymbol{\mu}$ and $ \hat{\boldsymbol{\mu}}_{MLE}$  is  $3.9^{\circ}$ only. In fact,
 using the Fisher distribution, it is found that $H_0$  is accepted with very high probability .
  Here $\hat{\kappa}_{MLE } =7.51$ so the data is not concentrated  and linear statistics  will not be appropriate. Further details are available, for example, in  \citet{mardiajupp2000}.

\begin{figure}[!htb]
\caption{\label{FisherColourOrg}Fisher 1953circa in a geological field trip in Australia; author's personal collection.}
\begin{center}
\includegraphics[width=15cm,angle=0]{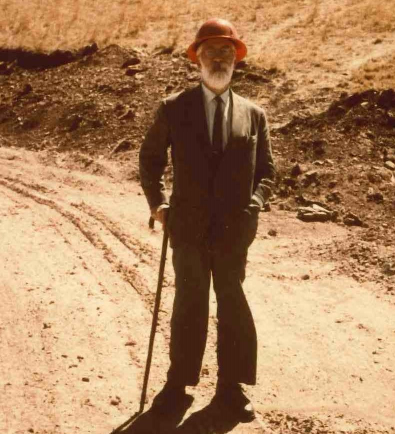}
\end{center}
\end{figure}

The topic continues to be of interest and \citet{watson1998} has given more details of  \citet{fisher1953} related to paleomagnetism   and continental  drift. C. R. Rao in  \citet{Rao2008} p.135 commented that
\\
 ``Fisher (1953) used his model to estimate the true direction ($\theta $) of remnant rock
magnetism in lava flow assuming that the observations collected over a geographical area
are independent. He did not consider the possibility of spatial correlations which may have
some effect on estimation.''
\\
 Very  recently \citet{scealy2022} have reassessed  \citet{fisher1953},  allowing for the site differences in Hospers type remanent magnetism data.
\\
We end this section with a  Fisher's photo (see,  Figure \ref{FisherColourOrg}) showing one of his  field trips for geological data. We quote \citet{box1978}, p.445, 
\\
``Fisher continued to the last to be fascinated by the geophysical explorations and to encourage and befriend the explorers.''
\\
 Note that his  geological explorations are not as  well known like  his genetic experiments (see, for example,  one of his genetic experiments on mice \citet{box1978}, pp. 379-381.)

\section{ Fisher on the von Mises distribution }\label {FisherOnVM}
Fisher  in his second edition of ``Statistical Methods and Scientific Inference”, 1959, p. 137,   goes back to support  his fiducial argument.  In so doing, he incidentally derives the  von Mises distribution  as follows.
Let  ${\bf  x}_1, \ldots, {\bf  x}_n,$ be  a random sample drawn from a bivariate normal distribution with the mean vector  ${\boldsymbol{\mu}}$, unit variance and zero correlation  so the density is given by 
$$
f({\bf  x}) = \exp\{- \frac{1}{2} ({\bf  x}-\boldsymbol{\mu} )^T ({\bf  x} -\boldsymbol{\mu} ) \}, \quad  {\bf  x},\boldsymbol{\mu}  \in \mathbb{R}^2.
$$
 Assume  ${\boldsymbol{\mu}}^T = (\rho\sin\theta, \rho \cos\theta)$  with $ \rho$  known;  the aim is to find the confidence interval (fiducial limits) for  $\boldsymbol{\mu}$. 
 The sample mean $\bar {\bf  x}$ is the minimal sufficient  statistics  for $\boldsymbol{\mu}$ and  the ancillary statistics is the resultant $ \bar R=\lVert \bar {\bf  x} \rVert$.
 Then he shows  that the distribution of $\hat \theta$   given   $ \bar R $ is the von Mises distribution 
$$
    f(\hat\theta |  \bar R )=\frac{\exp\{n \bar R \rho \cos(\hat \theta-\theta)\}}{2\pi I_0(n \bar R \rho )}.
$$
 He  seems not to be aware  of the  von Mises paper (\citet{vonM1918}.)
 \\
  There are two cases for the conditional distribution depending on $ \bar R$: 
  
   {\bf Case A:}  $ \bar R <\rho,$ and  {\bf Case B:} $ \bar R >\rho;$
\\   
and $ \bar R =\rho$  is a singular case.
 Suppose under  {\bf Case A} and  {\bf Case B}, we  have  observed  $\bar R_1 $ and $ \bar R_2 $  respectively.
Each case will give rise to a  different   $\hat \theta$ from the von Mises distribution;  we write

for  {\bf Case A:} $\hat \theta=\hat \theta _1$,  and for  {\bf Case B:} $\hat \theta=\hat \theta _2.$ 
\\
Now our inference can be based on $f(\hat\theta_1 |  \bar R_1 )$ and $ f(\hat\theta_2 |  \bar R_2 ).$  \citet{hinkley1980}  has provided more details.
  
\section{The two primary goals of Fisher's 1953 paper } \label{Goals}
We note from the  paper  \citet{fisher1953}, there are possible two primary goal of the paper.
\\
{\bf Goal 1.} ``To provide methodology for the analysis of more or less widely dispersed
measurements of direction such as frequently arise in geology.'' 
\\
and 
\\
{\bf Goal 2.} ``Finally, it is the opinion of the author that certain misapprehensions as to
 the nature of inductive inference  have arisen in examples drawn from the theory
of the normal distribution, by reason of the peculiar characteristics of that distribution,
and that the examination of these questions, in an analogous though
analytically different situation, will  exhibit them in a clearer light."

 The Fisher distribution  has
become a principal tool for analysing spherical data so {\bf Goal 1} is achieved  but  its role as a non-standard example
of fiducial inference has not received comparable attention so  the  assessment of  his  {\bf Goal 2} still continues, see, for example, \citet{bingham1980}.

 \section{ Wrapped tangent distributions} \label{wrap}
 For very concentrated data on a sphere, Gauss used a tangent projection, leading to linear statistics. We have pointed out, \citet{fisher1953}  introduced  his  spherical distribution when the data is not concentrated.
 
Another approach is to  use directional distributions by wrapping a multivariate distribution using an ``exponential map'' with a base point in the tangent space (see,  \citet{mardiajupp2000}, Section 13.4.2).
  In particular this gives rise to a  distribution on the sphere as an alternative to  the Fisher distribution; recently, this construction   was used for the  matrix Fisher type distribution in  \citet{Benton2023}, motivated by problems in machine learning on manifolds. We will  show that  their matrix  distribution has some serious limitation (though not  for a very concentrated data.) If $g(\bm x)$  is a pdf  in the tangent space, $\bm x$ in  $\mathbb{R}^{q}$,  then the  ``wrapped  tangent" pdf (using the exponential map on the sphere with base point at the north pole $(0,\ldots, 0,1)$) ),  of $\bm y = (\sin \theta \bm v, \cos \theta)$, where $0 \leq \theta \leq \pi$  is the colatitude.
and $\bm v$ is a unit vector in $\mathbb{R}^{q-1}$ is given by 
\begin{equation}\label{wnMain}
f(\bm y) = (1/\sin^{q-2} \theta) \sum_{k=0}^\infty
\{ r_{1,k}^{q-2} g(r_{1,k} \bm v) + r_{2,k}^{q-2} g(-r_{2,k} \bm v)\}.
\end{equation}
The density $f(\bm y)$ is  with respect to  the uniform measure on the sphere,
where $r_{1,k} = \theta+2\pi k$, $r_{2,k} = 2\pi(k+1)-\theta$.
Except for the term involving $r_{1,0}$ at $\theta=0$, all the
remaining terms have a singularity at $\theta=0$ and at $\theta=\pi$.
In particular, this wrapped distribution on the sphere cannot be unimodal.

\begin{figure}[!htb]

\includegraphics[scale=.35,angle=0]{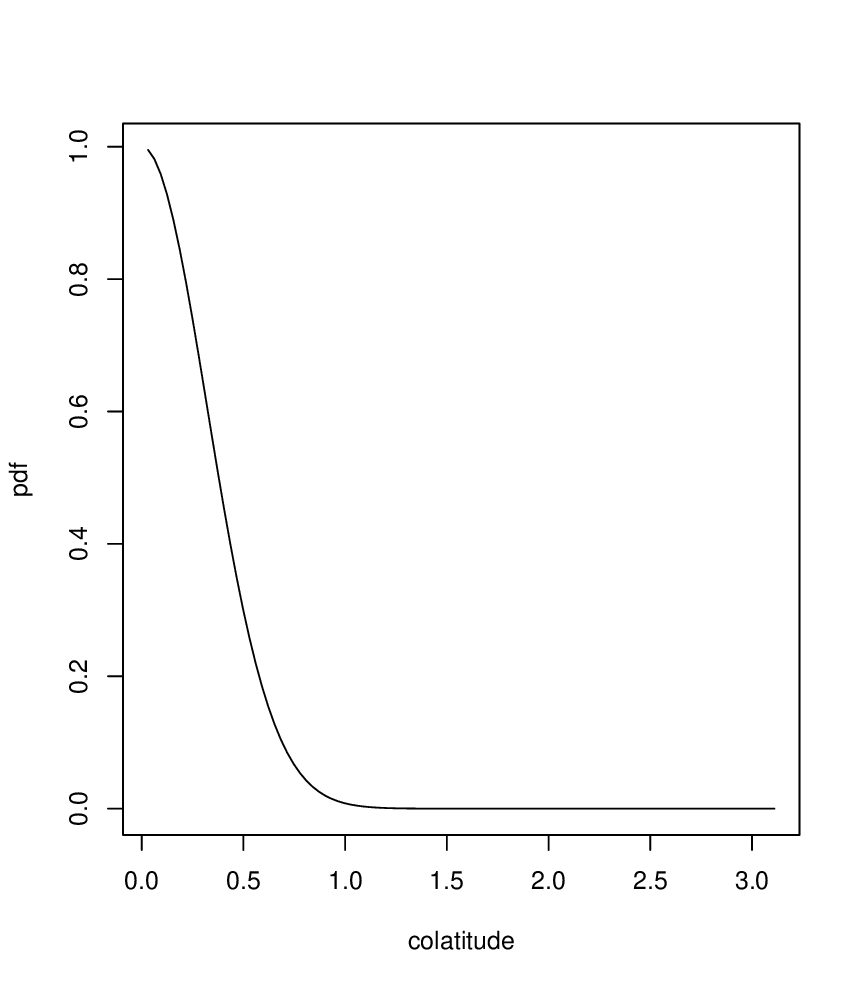}
\includegraphics[scale=.35,angle=0]{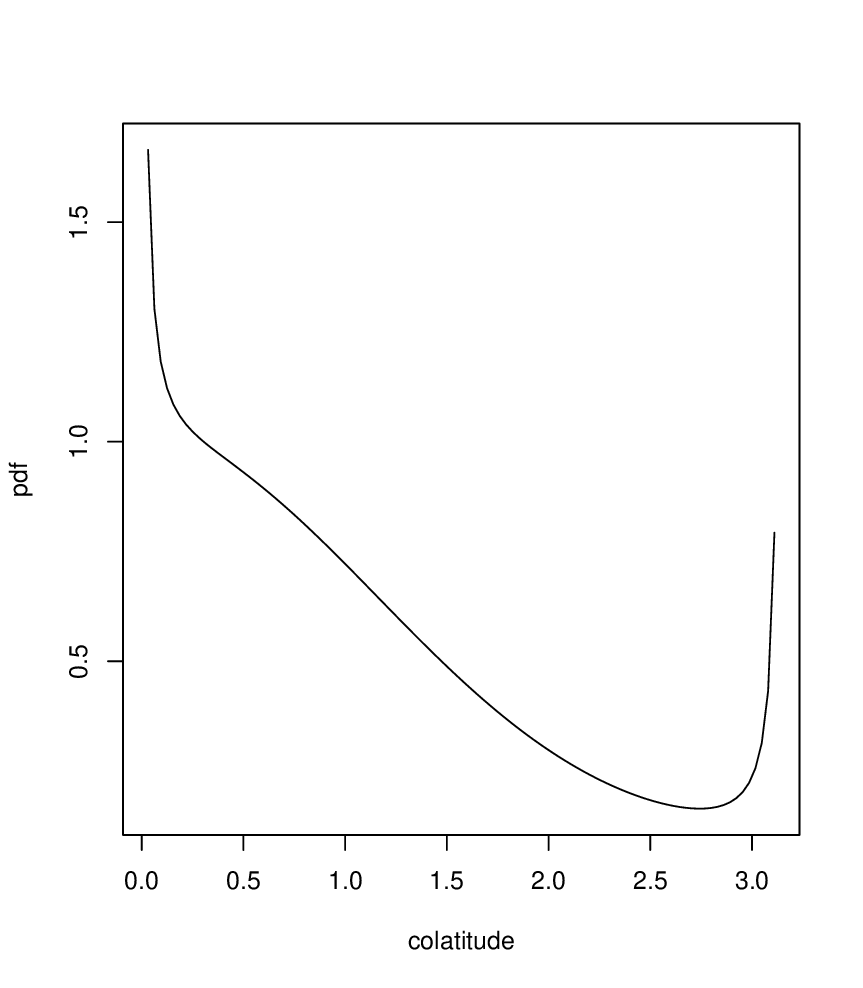}
 \includegraphics[scale=.35,angle=0]{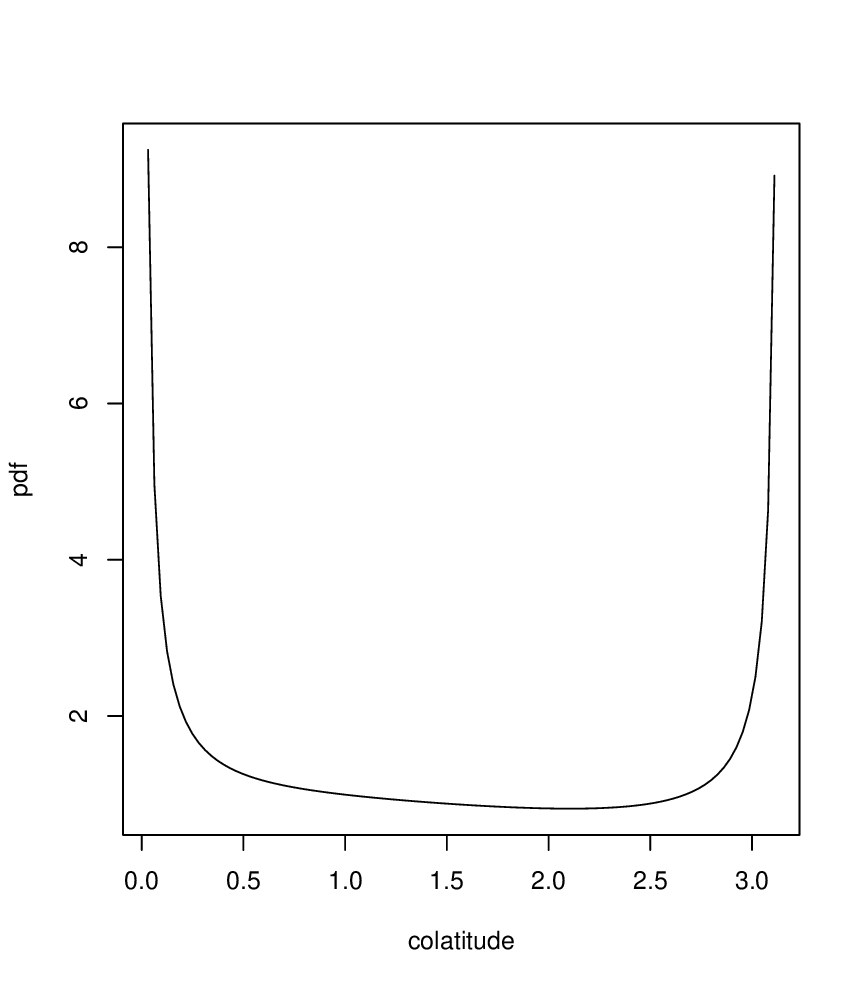}\\
 \caption{\label{pdfCo} The pdf of colatitude $\theta$  of the  wrapped tangent normal   distribution with 
   $\sigma^2=0.1, 1.0, \text {and}\; 2.0$  respectively.}
\end{figure}

When  $q=3$, with $f(\bm x)$  the normal distribution, we find that the pdf of 
the colatitude $\theta$ is 
\begin{equation}\label{wnSphere}
\frac {1}{\sin \theta} \sum_{k=0}^\infty
[ (\theta+2\pi k) \exp \{- (\theta+2\pi k)^2)/2\sigma^2\}+ (2\pi(k+1)-\theta) \exp \{-(2\pi(k+1)-\theta)^2/2\sigma^2\}],
\end{equation}
where $0 \leq \theta \leq \pi$.
Figure \ref{pdfCo} gives plots of the density $(\ref{wnSphere})$ for three values  $\sigma^2=0.1, 1.0, \text {and}\; 2.0$. As $\sigma>2$ increases the bi-modality  increases though for the concentrated data with  $\sigma^2=0.1$, it looks  unimodal though there are singularities; seen clearly for $\sigma^2=1.0, 2.0 .$
\\
\\
 For the circular case $q=2$,  the wrapped tangent pdf  can be written as
\begin{equation}\label{circ}
f(\theta) = \sum_{k=0}^\infty
\{ g(\theta+2\pi k ) +  g(2\pi(k+1)-\theta)\}
\end{equation}
so this reduces to the standard circular  wrapped distribution  except that we have  $0 \leq \theta \leq \pi$ as $\theta$ deemed to be ``colatitude'' for the circle. The standard representation by $\theta^*$ is the full circle so we have  $-\pi \leq \theta^* \leq \pi$. In the Euclidean representation 
$( \cos\theta^*,\sin\theta^*)$ is equivalent to $( \cos\theta,\pm\sin\theta)$.
\\
\\
Note that this  construction we have given here  is general:-
\\  
\\
Given a base point $m$ on the the Riemannian manifold and a vector $\bm x$ in the
tangent space, the exponential map $\exp_{m}(\bm x)$ yields a point on
the manifold and we can go from $g(.)$ to $f(.)$.
\\
\\
\citet{Benton2023} have used the wrapped normal on $SO(3)$ using the exponential  map to illustrate some of their work in Machine Learning.  However, the most common distribution used on $SO(3)$ is the Fisher matrix distribution which has many desirable properties (see, for example, \citet{mardiajupp2000}) whereas  
 there are again  inherent singularities in this wrapping   (see, also \citet{Mardia2023} for some more details). Since $SO(3)$ can be identified
with $S^3$ (after identifying antipodal points), any calculation on $SO(3)$ can be
reformulated as one on $S^3$ and the above discussion for  the spherical pdf  applies.
 This singularity of the pdf  
is not a practical  issue if $g(.)$ is highly concentrated near the origin
on the tangent plane, but it is an issue for more diffused
distributions and particularly can be an issue when used in  mixtures, see,  for example,  \citet{Mardia2023}.

\section {R. A. Fisher and   D'Arcy Thompson and Different Manifolds}\label{shape}

\begin{figure}[!htb]
\caption { \label{Darcy1}Letter from Fisher to D'Arcy  Thompson  in  1933.  }
\begin{center}
\includegraphics[scale=.95,angle=0]{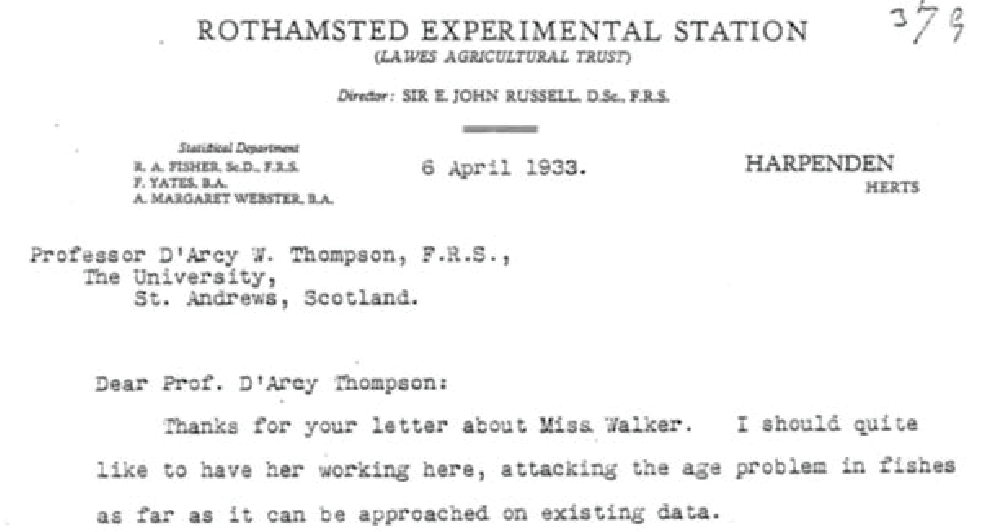}
\end{center}
\end{figure}

%%%%%%%%%%%
\begin{figure}[!htb]
\caption { \label{Darcy2} Letter (continued)  from Fisher to D'Arcy  Thompson  in  1933.   }
\begin{center}
\includegraphics[scale=.86,angle=0]{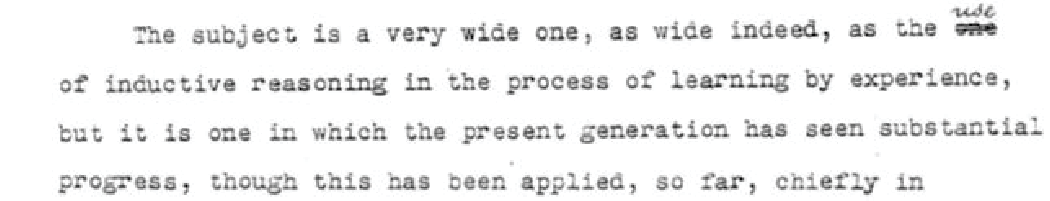} \;\;\;\;\;\;\;\;\;\;\;\;\;\;\;
\includegraphics[scale=.90,angle=0]{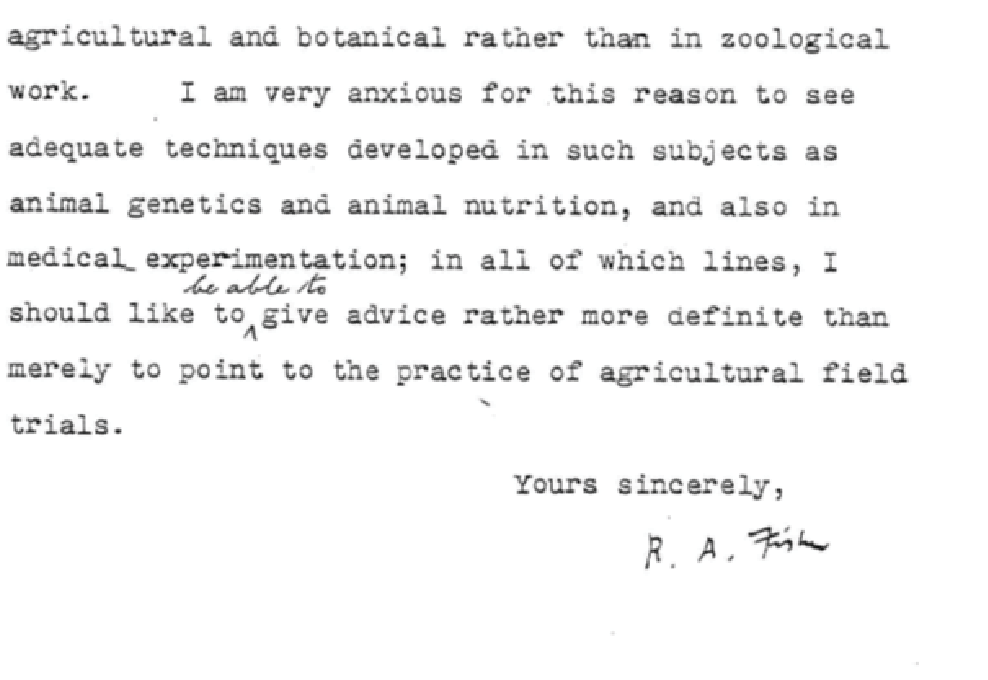}
\end{center}
\end{figure}

D'Arcy Thompson is the pioneer of Shape Analysis (see, for example, \citet{Dryden2016} p.2.); a subject which has a different manifold than what we have mentioned for directional statistics.
There seems to be no record of what Thompson wrote  to Fisher about  a student (Miss Walker)  but the letter Figure \ref{Darcy1} from Fisher   mentions her, as well as investigating     the age problem in fish. Fisher  wrote on 6 April 1933   to  Thompson  (see Figure \ref{Darcy2} ) that  the development of  adequate techniques as required in the area would be undertaken.  But sadly this  collaboration  via the student  did not happen.
%%%%%%

\section {Acknowledgments}  I wish to thank Kit Bingham, Colin Goodall, John Kent, Charles Taylor and Xiangyu Wu for their  helpful comments, and to the Fisher Memorial Trust for  inviting me to give the talk. Thanks are also due to the Leverhulme Trust for the Emeritus Fellowship.

\pagebreak
\bibliographystyle{rss}

%\bibliography{ref_PDWC}

\end{document}